\documentclass[prd,twocolumn,amsmath,amssymb,floatfix,nofootinbib,longbibliography,superscriptaddress]{revtex4-1}
\pdfoutput=1
\usepackage[T1]{fontenc}
\usepackage{amsthm}
\usepackage{amsmath}
\usepackage{amssymb}
\usepackage{graphicx}
\usepackage{natbib,ifthen}
\usepackage{color}

\usepackage{hyperref}

\usepackage{breakurl}

\makeatletter

\providecommand{\tabularnewline}{\\}
\newcommand{\rhomax}{\rho_{\rm max}}
\newcommand{\half}{\frac{1}{2}}

\makeatother

\begin{document}

\title{Cosmology with Negative Absolute Temperatures}

\author{J. P. P. Vieira}
\affiliation{Department of Physics \& Astronomy, University of Sussex, Brighton BN1 9QH, UK}
\email{J.Pinto-Vieira@sussex.ac.uk}

\author{Christian T. Byrnes}
\affiliation{Department of Physics \& Astronomy, University of Sussex, Brighton BN1 9QH, UK}

\author{Antony Lewis}
\affiliation{Department of Physics \& Astronomy, University of Sussex, Brighton BN1 9QH, UK}

\begin{abstract}

Negative absolute temperatures (NAT) are an exotic thermodynamical
consequence of quantum physics which has been known since the 1950's
(having been achieved in the lab on a number of occasions). Recently,
the work of Braun et al \cite{BraunNAT} has rekindled interest in negative temperatures
and hinted at a possibility of using NAT systems in the lab as dark
energy analogues. This paper goes one step further, looking into the cosmological consequences
of the existence of a NAT component in the Universe. NAT-dominated expanding Universes experience a borderline phantom expansion ($w<-1$) with no Big Rip, and their contracting counterparts are
forced to bounce after the energy density becomes sufficiently large.
Both scenarios might be used to solve horizon and
flatness problems analogously to standard inflation and bouncing cosmologies.
We discuss the difficulties in obtaining and ending a NAT-dominated epoch, and
possible ways of obtaining density perturbations with an acceptable spectrum.

\end{abstract}

\maketitle

\tableofcontents{}

\pagebreak{}

\section{Introduction}

\subsection{How can temperature be negative?}

Say the words ``negative absolute temperatures'' (NAT) to anyone
who hasn't heard of them before, and your remark will most likely be
met with a look of bewilderment (and perhaps the question in the title).
Even more than sixty years after negative temperatures were achieved in the lab, this
is by no means an unexpected reaction. In informal parlance we all get used to
perceiving temperature as a measure of the energy in a macroscopic system, and thus necessarily a positive
quantity.
In fact, temperature is canonically defined in terms of the rate of change
of entropy with internal energy in thermal equilibrium, which can be negative. Specifically
\begin{equation}
\frac{1}{T}=\left(\frac{\partial S}{\partial U}\right)_{V,\, N,\, X_{i}}\label{eq:thermo_T}
\end{equation}
where $T$ is the absolute temperature, $U$ the internal energy,
$S$ the entropy, $V$ the volume, $N$ the number of particles and
$X_{i}$ represents any other (eventually) relevant extensive property
of the system. In this work, $S$ is defined as%
\footnote{
\label{fn:Boltazmann-Gibbs_controversy}
There has recently been some controversy \cite{inconsist_NAT,comment_consist_NAT,Gibbs_Boltzmann_Teq,Repy_comment_inconsist_NAT,Boltz_vs_Gibbs,thermo_isolated,inform_gibbshertz,dispute_boltzgibbs}
regarding whether this quantity, known as the Boltzmann entropy, is correct;
the alternative being the Gibbs-Hertz entropy, brought under the spotlight
by \cite{inconsist_NAT} (in the original reference, they just call
it the ``Gibbs entropy'' since Gibbs was apparently the first to
propose this entropy formula despite it traditionally being credited
to Hertz). While this debate is an important one (especially for
anyone interested in NAT, which are impossible in the Gibbs-Hertz
formalism), it is not completely clear in which situations the disagreement
actually affects obervables in the thermodynamic limit \cite{Boltz_vs_Gibbs}.
Moreover, it has recently been shown \cite{dispute_boltzgibbs}
that the Boltzmann formula is the appropriate one for systems
with equivalence of statistical ensembles.
}
\begin{equation}
S=k_{B}\ln W\label{eq:Boltzmann_entropy}
\end{equation}
where $k_{B}$ is the Boltzmann constant and $W$ is the number
of microstates corresponding to the macrostate the system is in.

The reason we do not expect NAT in classical scenarios is that for
those we generally expect the number of states with energy $U$ to
increase with $U$. In quantum mechanical systems, however, it is
fairly easy to construct situations in which the energy is bounded
from above as well as from below. When that happens, if the entropy
is a continuous function of the energy, $S$ must
have a maximum somewhere between the upper and the lower energy bounds
(where $S$ is zero). By Eq.~\ref{eq:thermo_T}, $T$ must then
allow negative values.

The simplest example is a two-level quantum
system which can be populated by a fixed number of distinguishable
particles. As the energy of the system is increased, more particles will populate the higher-energy level.
At infinite temperature the number of particles is the same in both energy levels
(corresponding to maximum entropy), but it
is quite possible to give the system more energy than that, so that there are then more particles in the
higher-energy state, corresponding to a negative temperature. Note that the system at a negative temperature
has more energy, and is therefore ``hotter'', than at a positive temperature.

In practice, negative temperatures can be realized in a number of ways. As an illustration,
consider a lattice of localized spin-$1/2$ particles interacting with an external
magnetic field. There are two one-particle energy levels, corresponding
to the two possible spin orientations relative to the magnetic field.
At low temperatures, we expect most spins to be in the lowest-energy
state. However, if the sign of the external magnetic field is switched
at very low temperatures, then suddenly the most populated state will
become the highest-energy state and if the system can then be isolated
(so that energy cannot be lost and most particles are forced to be
in the highest-energy state) then we are left with a state corresponding
to $T<0$. This was essentially the set-up used by Purcell and Pound
in 1951 \cite{NuclearNAT}, in the first experiment in which it is
claimed that NAT were measured (the magnetic material they used was
crystal of Lithium fluoride, which was known to have very long magnetic relaxation
times).
% and is still considered the archetype of a system that admits
%NAT.

\subsection{From the lab to the sky}

The first thorough theoretical study of the conditions under which
NAT occur is due to Ramsay \cite{RamseyNAT}, five years after the
experiment by Purcell and Pound \cite{NuclearNAT} (although the first
known appearance of the concept of NAT seems to have been two years earlier,
when Onsager used them to explain the formation of large-scale
persistent vortices in turbulent flows \cite{NAT_turbul_vortex}). Even today, most
discussions revolving around NAT take this treatise as a starting
point.

After Ramsay (1956), there was not much big news regarding NAT until
2012, when Braun et al.~\cite{BraunNAT} reported the first experimental
realization of NAT in a system with motional degrees of freedom (an ultra-cold
boson gas). Important as this may be as an experimental landmark,
one of its main consequences was arguably the revival of theoretical
interest in NAT which led to the debate about Boltzmann vs Gibbs-Hertz
entropies we have already mentioned (see footnote \ref{fn:Boltazmann-Gibbs_controversy}).
%which raises important questions about thermodynamics in small systems and
%the nature (and sign) of temperature.
Interestingly, Braun et al. also noticed that an (almost) inevitable consequence
of negative temperatures, negative pressures, are \emph{``of fundamental
interest to the description of dark energy in cosmology, where negative
pressure is required to account for the accelerating expansion of
the universe''}. Apparently, this remark was mostly interpreted as a
suggestion that known NAT systems could be useful as laboratory dark
energy analogues. Some people, however, seem to have read this hint
differently, meaning that some analogous mechanism could be responsible
for the observed accelerated expansion of the Universe. This interpretation
seems to have inspired Brevik and Gr\o n \cite{NEGVISC} to come up
with a class of models where, while not using NAT directly, an analogous
effect is achieved by means of a negative bulk viscosity. Nevertheless,
as far as we are aware, nobody has proposed a model where this is
done with actual negative temperatures, possibly due to not having
found a well-motivated physical assumption that could lead to NAT
at cosmological scales%
\footnote{A connection between NAT and phantom inflation seems to have
been first independently suggested in Ref.~\cite{Phantom_thermo}. However,
the word "temperature" there is really referring to an out-of-equilibrium
generalisation of temperature and none of their examples can
correspond to NAT as defined here. Those following the ensuing
discussion \cite{thermo_phantom2,thermo_phantom_mu1,thermo_phantom_mu2,thermo_phantom_mu3,thermo_phantom_mu4,scalarNATmail}
might be interested in the questions we raise regarding the introduction of a non-null
chemical potential in this context (see Appendix~\ref{sec:The-problem-of-mu}).%
}.

\subsection{A natural cut-off?\label{sub:A-natural-cut-off?}}

The key requirement for a NAT is an upper bound to the energy of the system.
This could either be an absolute upper bound, or there could be an energy gap allowing a metastable
population inversion. As long as the interaction time for particles below the energy gap is much shorter
than the typical time scale for thermal equilibrium to be reached,
an effective NAT can develop (as in the experimental realizations).

In the context of cosmology, where we are mainly interested in the properties of the total density,
a NAT could be obtained if there is a fundamental energy cut-off. This could
 be related to a minimum length scale, for example as discussed
in the context of quantum gravity (see for example \cite{min_length_review,Garay_min_length}
and references therein). For the purpose of this paper we are not assuming any particular model,
 and simply consider the possibility that the fundamental model features a cutoff and investigate the consequences.
  Since the NAT description also requires thermal equilibrium, we also require the interaction
time for dominant particles with energies up to the cut-off to be short compared to other relevant timescales.
Of course any population inversion could instead rapidly go out of equilibrium as the particles decouple, but
we focus on the possibility that equilibrium is maintained and see what a phenomenological NAT description would imply.

The relevant quantity that needs to
be extracted from an eventual fundamental theory is the number density
of states at a given energy $\epsilon$, $g\left(\epsilon\right)$,
which at low energies is constrained to take a standard form.
Given the lack of an actual complete fundamental theory to
work with, we shall express all results in the most general form possible.
Any time we want to illustrate a calculation for a specific model we consider
a simple ansatz for a gas of independent particles with a cut-off at $\epsilon=\Lambda$
and the right behaviour at low (i.e., currently observed) energies,
\begin{equation}
g\left(\epsilon\right)=\begin{cases}
\frac{g}{2\pi^{2}}\epsilon\sqrt{\epsilon^{2}-m^{2}} & \text{if}\ m<\epsilon<\Lambda\\
0 & \text{otherwise},
\end{cases}\label{eq:DOS_Lambda}
\end{equation}
where $g$ is the usual degeneracy factor and $m$ is the particle
mass (note we are using units in which $c=\hbar=M_{P}=k_{B}=1$).
Interestingly, it turns out that our most important results
in section \ref{sec:NATive-Cosmology} will be essentially independent
of the specific form of $g\left(\epsilon\right)$ as long as it behaves as it should at low energies.

In the remainder of this paper, we shall focus on the cosmological
implications of NAT. The discussion is organised as follows.
In section \ref{sec:Thermodynamical-functions} we show how to calculate
thermodynamical functions as model-independently as possible. In section
\ref{sec:NATive-Cosmology} we use the results from section \ref{sec:Thermodynamical-functions}
to model the evolution of generic expanding and contracting NATive
Universes. In particular, we show that exactly exponential inflation
is an attractor regime in these models and address the problems
associated with ending it. Finally, the main successes and problems
of this approach are summarised in section \ref{sec:Conclusions}.
Additionally, appendix \ref{sec:Thermal-Perturbation-Generation}
deals with the challenges of thermal perturbation generation at NAT.

\section{Thermodynamical functions\label{sec:Thermodynamical-functions}}

The main goal of this section is to investigate the temperature dependence
of the most relevant thermodynamical quantities (which we will later
need to substitute into the Friedmann equations in order to do cosmology).
In particular, we are interested in finding model-independent relations
between results at very low positive temperatures (the kind that has
been extensively studied) and results at negative temperatures very
close to $T=0^{-}$ (which we shall see generally corresponds to the
highest possible energy scales, which have never been probed).

\subsection{Negative pressure}

Our main motivation for studying NAT is that they naturally
give rise to negative pressures. Let us start by seeing why this is so.
One of the most straightforward ways of calculating the pressure of
a system is by making use of the grand potential, defined as
\begin{equation}
\Phi=U-ST-\mu N, \label{eq:Helmholtz}
\end{equation}
and whose gradient can be written as
\begin{equation}
d\Phi=-SdT-PdV+N d\mu+x^{i}dX_{i}, \label{eq:dF}
\end{equation}
where $\mu$ is the chemical potential and $x^{i}$ represent the
thermodynamic potentials corresponding to the quantities $X_{i}$.
Assuming there is no relevant $X_{i}$, we get the well-known Euler
relation:
\begin{equation}
P=-\left(\frac{\partial \Phi}{\partial V}\right)_{T,\, \mu}=-\rho+sT+\mu n.\label{eq:P_general}
\end{equation}

Note that when $T<0$ the only term in Eq.~\ref{eq:P_general} which
is not necessarily negative is $\mu n$, and the pressure will be
very negative unless this term is significant in comparison to the
others. In particular, if $\mu=0$ (as must be the case in regimes
where the total number is not conserved) we recover the better-known
result
\begin{equation}
P=-\rho+sT,\label{eq:P_mu0}
\end{equation}
which corresponds to an equation of state with $w<-1$ (leading to
what is known as \emph{phantom inflation}) for any $T<0$.

\subsection{Fermions and holes}

For now we deal only with fermions (in appendix
\ref{sec:The-problem-of-mu} we discuss why we do not want to work
with bosons). We will therefore use the
Fermi-Dirac distribution,
\begin{equation}
{\cal N} \left(\epsilon;T,\mu\right)=\frac{1}{e^{\beta\left(\epsilon-\mu\right)}+1}, \label{eq:fermi_dirac}
\end{equation}
which should still be valid for $\beta=\left(k_{B}T\right)^{-1}<0$
since microstate probabilities are still associated with the Boltzmann
factor $e^{-\beta\left(E-\mu N\right)}$ (where $E$ is the total
energy associated with a specific microstate, so that $U=\left\langle E\right\rangle $).

We can now use standard thermostatistics to find the relevant quantities
as a function of temperature and chemical potential.
The energy and the number density are trivial,
\begin{equation}
\rho\left(T,\mu\right)=\intop_{m}^{\Lambda}\epsilon g\left(\epsilon\right){\cal N}\left(\epsilon;T,\mu\right)d\epsilon, \label{eq:rho_general}
\end{equation}
\begin{equation}
n\left(T,\mu\right)=\intop_{m}^{\Lambda}g\left(\epsilon\right){\cal N}\left(\epsilon;T,\mu\right)d\epsilon, \label{eq:n_general}
\end{equation}
as are their maximum possible values,
\begin{equation}
\rho_{\mathrm{max}}\equiv\intop_{m}^{\Lambda}\epsilon g\left(\epsilon\right)d\epsilon, \label{eq:rhomax}
\end{equation}
\begin{equation}
n_{\mathrm{max}}\equiv\intop_{m}^{\Lambda}g\left(\epsilon\right)d\epsilon.\label{eq:nmax}
\end{equation}
Note that these maximum values correspond only to the NAT fermion gas, so
in situations in which there is more than one component the total $\rho$ and $n$
can exceed these values.

The pressure is less simple, but can be found from the grand potential given by \cite{blundell}
\begin{equation}
\Phi=-\frac{1}{\beta}\ln Z, \label{eq:F_Z}
\end{equation}
where $Z$ is the grand canonical partition function. For fermions this
is just given by
\begin{multline}
Z=\sum_{s}e^{-\beta\left(E_{s}-\mu N_{s}\right)}=\sum_{\left\{ N_{i}\right\} }\prod_{i}e^{-\beta\left(\epsilon_{i}-\mu\right)N_{i}}
\\
=\prod_{i}\left(1+e^{-\beta\left(\epsilon_{i}-\mu\right)}\right), \label{eq:Zg_fermions}
\end{multline}
where $s$ are the states of the whole system and we used $i$ to
label different one-particle states, $\epsilon_{i}$ and $N_{i}$
representing their energy and occupation number ($0$ or $1$) respectively,
and $\left\{ N_{i}\right\} $ represents a sum over all possible combinations
of $N_{i}$. Inserting Eq.~\ref{eq:Zg_fermions} into Eq.~\ref{eq:F_Z}
and then taking the continuous limit before applying Eq.~\ref{eq:P_general} we finally find
\begin{equation}
P\left(T,\mu\right)=\frac{1}{\beta}\intop_{m}^{\Lambda}g\left(\epsilon\right)\ln\left[1+e^{-\beta\left(\epsilon-\mu\right)}\right]d\epsilon.\label{eq:f(T,mu)}
\end{equation}

So far, it looks as though all these results should be highly dependent
on the specific form of $g\left(\epsilon\right)$. The reason this
is not true is because we can relate results at positive and negative
temperatures using the well-known symmetry of the Fermi-Dirac distribution:
\begin{multline}
{\cal N}\left(\epsilon;T,\mu\right)=\frac{1}{e^{\beta\left(\epsilon-\mu\right)}+1}=1-\frac{1}{e^{-\beta\left(\epsilon-\mu\right)}+1}
\\
=1-{\cal N}\left(\epsilon;-T,\mu\right).\label{eq:hole_sym}
\end{multline}
This allows us to borrow the concept of \emph{holes} from solid state
physics.
A hole here is just a way to conceptualize the absence of
a particle in a given state as a quasi-particle of negative energy
in a positive energy ``vacuum''. This just means that it is as valid
to describe our system in terms of which states are occupied by particles
as in terms of which states are unoccupied. For us it is particularly
useful in the limit where most particles are occupying the highest-energy
states (which correspond to $T<0$), since this can be thought of
as the limit where holes are populating the lower-energy states (corresponding
to $T>0$). Note that there exists a similar identity for the kind
of logarithmic term in the integral in Eq.~\ref{eq:f(T,mu)},
\begin{equation}
\ln\left[1+e^{-\beta\left(\epsilon-\mu\right)}\right]=-\beta\left(\epsilon-\mu\right)+\ln\left[1+e^{\beta\left(\epsilon-\mu\right)}\right].\label{eq:log_property}
\end{equation}

It is now easy to combine Eqs.~\ref{eq:hole_sym} and \ref{eq:log_property}
with Eqs.~\ref{eq:rho_general}, \ref{eq:n_general}, and \ref{eq:f(T,mu)}, yielding

\begin{equation}
\rho\left(T,\mu\right)=\rho_{\mathrm{max}}-\rho\left(-T,\mu\right)\label{eq:rho_holes}
\end{equation}
\begin{equation}
n\left(T,\mu\right)=n_{\mathrm{max}}-n\left(-T,\mu\right)\label{eq:n_holes}
\end{equation}
\begin{equation}
P\left(T,\mu\right)=-\rho_{\mathrm{max}}+\mu n_{\mathrm{max}}-P\left(-T,\mu\right).\label{eq:P_holes}
\end{equation}
These functions depend on very few
parameters from the fundamental theory as long as holes are at ``low''
temperatures (which here just means low enough that we know how physics
works at those temperatures). If $\mu=0$, as will be the case in
most relevant scenarios in this paper, the pressure has an even simpler
form%
\footnote{It is interesting to notice that this seemingly surprising relation
still makes sense physically. Since (if $\mu=0$) $P=-\left(\frac{\partial U}{\partial V}\right)_{S}$,
in a situation in which all single-particle states are filled the
entropy is zero, and keeping the entropy constant as $V$ varies corresponds
to always keeping all states filled, yielding $U=\rho_{\mathrm{max}}V$ and
thus $P=-\rho_{\mathrm{max}}$. If not all states are filled, then it makes
sense to think of holes as negative momentum particles that contribute
negatively to the total pressure as in Eq.~\ref{eq:P_holes_mu0}.%
}:
\begin{equation}
P\left(T,\mu=0\right)=-\rho_{\mathrm{max}}-P\left(-T,\mu=0\right)\label{eq:P_holes_mu0}
\end{equation}
(note that only in this case can we be sure that a barotropic fluid
at $T>0$ will correspond to a barotropic fluid at $T<0$). Note also
the useful symmetry
\begin{multline}
\rho\left(T,\mu=0\right)+P\left(T,\mu=0\right)=
\\
-\rho\left(-T,\mu=0\right)-P\left(-T,\mu=0\right).\label{eq:rho+P_sym}
\end{multline}

If, in addition to having $\mu=0$ and $T<0$, we have holes behaving
like cold matter (corresponding to $m\gg-T$), the quantity $\rho + P$ and the equation of state parameter $w\equiv P/\rho$
are given by
\begin{equation}
\rho + P= \rho -\rho_{\rm{max}}<0,\ \ \ \ \ \ w=-\frac{\rho_{\mathrm{max}}}{\rho}<-1,\label{eq:w_mu0_matter}
\end{equation}
whereas if they behave like radiation (the opposite limit)
\begin{equation}
\rho + P= \frac{4}{3}\left(\rho -\rho_{\rm{max}}\right)<0,\ \ \ \ \ \ w=-\frac{1}{3}\left(4\frac{\rho_{\mathrm{max}}}{\rho}-1\right)<-1.\label{eq:w_mu0_radiation}
\end{equation}

Alternatively, it can be interesting to consider the high $|T|$ region separating $T<0$ and $T>0$,
where $\left|\beta\Lambda\right|\ll1$.
Then, just looking at the limit when $\beta\rightarrow0$ yields
(from Eqs.~\ref{eq:rho_general} and \ref{eq:f(T,mu)}),
to leading order in $\beta$ and still assuming $\mu=0$,
\begin{equation}
\begin{cases}
\rho=\frac{1}{2}\rho_{\mathrm{max}}-\frac{\left\langle \epsilon^{2}\right\rangle _{0}}{4}\beta\\
P=\frac{\ln2}{\beta}n_{\mathrm{max}}-\frac{1}{2}\rho_{\mathrm{max}}+\frac{\left\langle \epsilon^{2}\right\rangle _{0}}{8}\beta
\end{cases},\label{eq:mu_T_inf}
\end{equation}
where
\begin{equation}
\left\langle \epsilon^{n}\right\rangle _{0}\equiv\intop_{m}^{\Lambda}\epsilon^{n}g\left(\epsilon\right)d\epsilon.\label{eq:avg_eps}
\end{equation}

\begin{figure}
\noindent \begin{centering}
\includegraphics[scale=0.35]{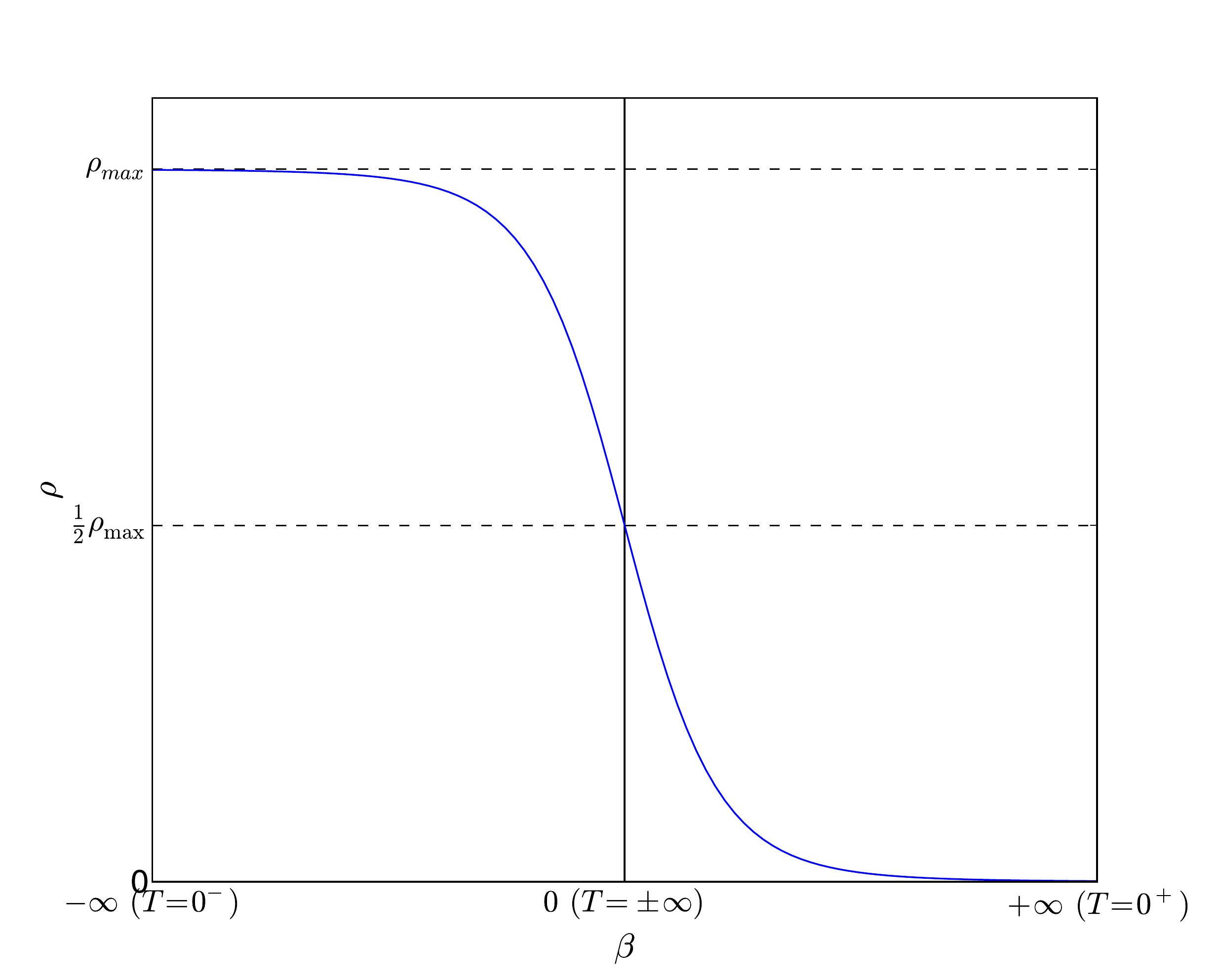}
\par\end{centering}

\noindent \begin{centering}
\includegraphics[scale=0.35]{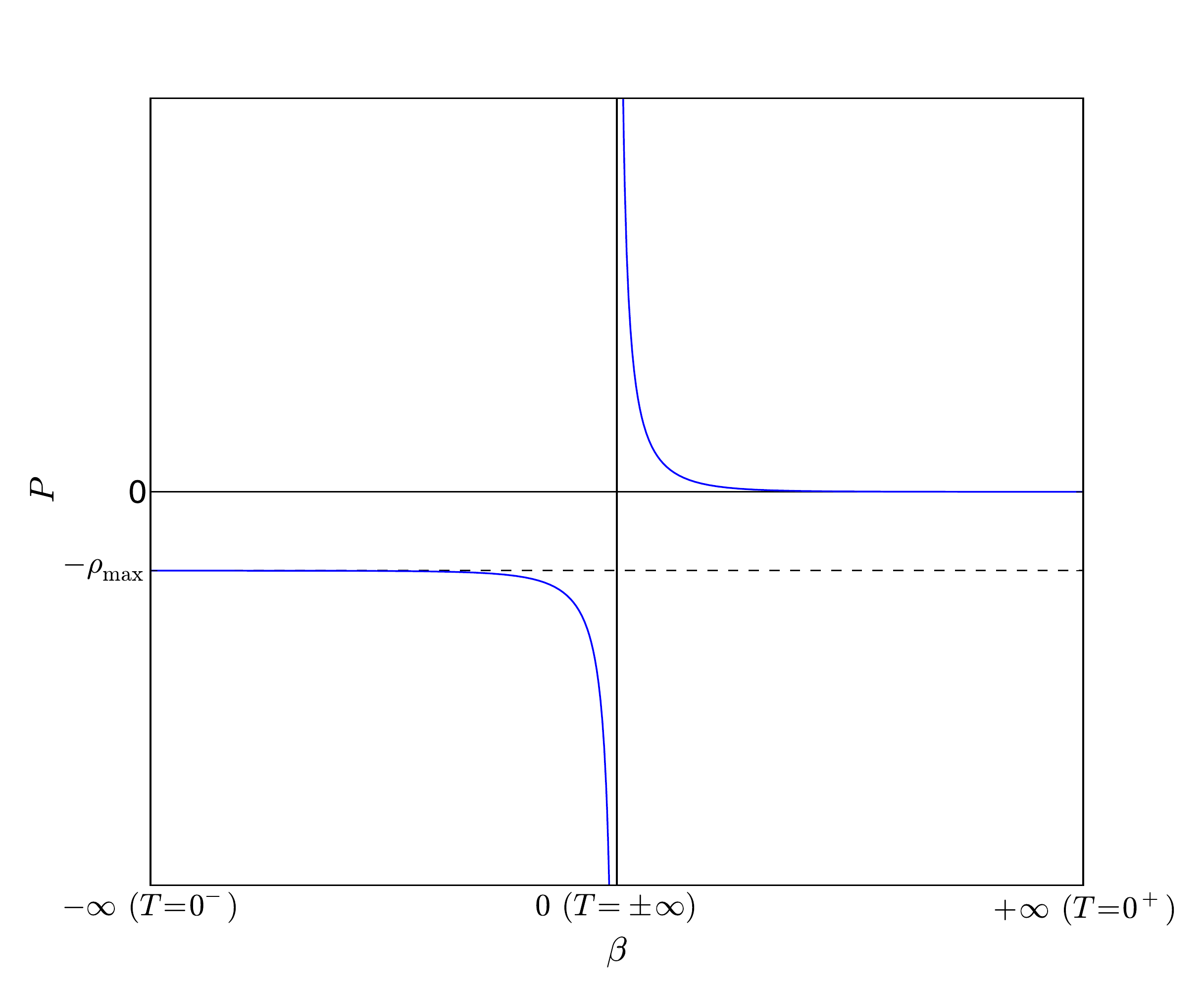}
\par\end{centering}

\noindent \raggedright{}\caption{\label{fig:Energy-density-and}Energy density and pressure as functions
of $\beta$ for a massless fermion with $\mu=0$ and $g\left(\epsilon\right)$
given by Eq.~\ref{eq:DOS_Lambda}.
}
\end{figure}

Notice that thanks to this we can know that the energy density and pressure profiles
have to look like those in Fig.~\ref{fig:Energy-density-and} (except for intermediate
values of $\beta$).

\section{Cosmology\label{sec:NATive-Cosmology}}

We are finally ready to investigate the cosmological consequences
of NAT. In this section we answer the question ``How does a Universe
at negative absolute temperature behave?''. In order to answer
this, and motivated by our analysis so far, we first assume that a FLRW
Universe is filled by a single perfect fluid in thermal equilibrium at NAT,
 and that this fluid is made up of fermions not subject to number
conservation (which we shall refer to as \emph{temperons}).
The requirement of thermal equilibrium can probably be translated
into a requirement for temperon-producing interactions to operate quickly compared to the Hubble time.
We do not consider scenarios with number conservation and/or bosons
because those entail additional (model-dependent) problems discussed in Appendix \ref{sec:The-problem-of-mu}\footnote{Note that, even if those problems can be overcome, the cosmological relevance of temperons subject to number conservation is reduced by the fact that they cannot play an important role in the dynamics of an expanding Universe for more than a few e-foldings due to their quick dilution (although they might play a role in a contracting or bouncing scenario).}. We further assume that at ``low'' energy scales
these temperons should behave like all other known particles; i.e.,
like matter or radiation, depending on their mass.

New physics giving rise to the maximum energy cutoff required for NAT could produce new dynamics when many particles have energies close to the cutoff.
However, to make progress, here we simply assume that general relativity still holds at the relevant macroscopic scales so that
the dynamics of the NATive Universe will then be governed by the usual Friedmann equations
\begin{equation}
\begin{cases}
3H^{2}=\rho\\
\dot{H}=-\frac{1}{2}\left(\rho+P\right),
\end{cases}\label{eq:Friedmann_eq_Lambda0_k0}
\end{equation}
where $\rho$ and $P$ will be calculated according to the process
outlined in section \ref{sec:Thermodynamical-functions}. The energy conservation equation,
\begin{equation}
\dot{\rho}=-3H\left(\rho + P\right),\label{eq:rhocons}
\end{equation}
can also be integrated to give a useful relation between the number of e-foldings the Universe has
expanded (or contracted) and its initial and final energy densities:
\begin{equation}
N=-\frac{1}{3}\intop_{\rho_{i}}^{\rho_{f}}\frac{d\rho}{\rho+P}=-\frac{1}{3}\intop_{\rho_{i}}^{\rho_{f}}\frac{d\rho/\rho}{1+w},\label{eq:DN_rho}
\end{equation}
where, as usual, $N\equiv\ln\frac{a_{f}}{a_{i}}$ and subscripts $i$
and $f$ denote ``initial'' and ``final'', respectively.

In the first two subsections of this section we shall focus on analysing
the background dynamics of two qualitatively different scenarios:
NAT in expanding cosmologies (subsection \ref{sub:NATive-inflation}),
and NAT in contracting cosmologies (subsection \ref{sub:NATive-bounces}).
% and argue that both provide a possible way of solving horizon and
%latness problems.
The remainder of this section will then be dedicated
to discussing perturbation generation and the transition to a normal positive-temperature
Universe.

\subsection{NATive inflation\label{sub:NATive-inflation}}

It can be easily seen from Eq.~\ref{eq:rhocons} that expanding cosmological solutions with negative temperature ($H>0$, $T<0$) have an attractor fixed point
at $T=0^{-}$, corresponding to de Sitter expansion with $\rho=\rho_{\mathrm{max}}$ and $w=-1$.
This has the interesting consequence that
all expanding NATive Universes should tend towards a phase of exactly
exponential inflation (although not necessarily reaching it)%
\footnote{This property suggests it might be possible to explain the
accelerated expansion we measure today with a \emph{dark temperon}
component. Unfortunately, any such mechanism would have to rely on
a very low energy cut-off, and one would have to explain why this
dark temperon behaves so differently from every other particle at
that energy (we would expect $\rho_{\mathrm{max}}\sim\rho_{\Lambda0}\simeq10.5h^{2}\Omega_{\Lambda0}\mathrm{GeVm^{-3}}$).%
} --- therefore, if this limit is reached, we should expect $\rho_{\rm{max}}\lesssim\rm{10^{111}GeVm^{-3}}$
just from the fact that we have not seen primordial tensor modes.

Interestingly, unlike with most phantom inflation models (recall that
Eq.~\ref{eq:P_mu0} implies our expansion must either be phantom or
exactly exponential), we do not have to worry about a \emph{Big Rip}
--- a divergence of the scale factor in a finite interval of time \cite{BigRip}.
This is simply because the energy density (and therefore $H$) here is
bounded, so the evolution asymptotes to exponential expansion with constant density sufficiently quickly
that the impact of the transient phantom period is small.

We start our quantitative analysis by showing that even if we begin
very close to $T=-\infty$ we should expect to evolve towards the
vicinity of $T=0^{-}$ extremely rapidly. If we are in the high $|T|$ regime where $|\beta\Lambda|\ll 1$ then, from
Eqs.~\ref{eq:mu_T_inf} and \ref{eq:DN_rho}, the number of e-foldings
between two densities while in this regime is
\begin{multline}
N\approx\frac{4/3}{\left(\ln 2\right)  n_{\mathrm{max}}\left\langle \epsilon^{2}\right\rangle _{0}}\intop_{\rho_{i}}^{\rho_{f}}\left(\rho-\frac{1}{2}\rho_{\mathrm{max}}\right)d\rho
\\
=\frac{2/3}{\left(\ln 2\right) n_{\mathrm{max}}\left\langle \epsilon^{2}\right\rangle _{0}}\left[\left(\rho_{f}-\frac{1}{2}\rho_{\mathrm{max}}\right)^{2}-\left(\rho_{i}-\frac{1}{2}\rho_{\mathrm{max}}\right)^{2}\right]
\\
\approx \frac{\left\langle \epsilon^{2}\right\rangle _{0}}{n_{\mathrm{max}}}\frac{\beta_{f}^{2}-\beta_{i}^{2}}{24\ln 2},\label{eq:DN_T_inf}
\end{multline}
where we have used
\begin{equation}
\beta\simeq\frac{4}{\left\langle \epsilon^{2}\right\rangle _{0}}\left(\frac{1}{2}\rho_{\mathrm{max}}-\rho\right).\label{eq:beta_zero_rho}
\end{equation}

In order to get some intuition regarding the order of magnitude we
should expect from this $N$, we can assume the simple ansatz from
Eq.~\ref{eq:DOS_Lambda} with $m=0$ (the order of magnitude should
not change significantly as long as $m\ll\Lambda$) and find
\begin{equation}
\begin{array}{c}
n_{\mathrm{max}}=\frac{g}{6\pi^{2}}\Lambda^{3}\\
\rho_{\mathrm{max}}=\frac{g}{8\pi^{2}}\Lambda^{4}\\
\left\langle \epsilon^{2}\right\rangle _{0}=\frac{g}{10\pi^{2}}\Lambda^{5}
\end{array}\label{eq:n_rho_eps2_max}
\end{equation}
leading to
\begin{equation}
N=\mathcal{O}\left\{ \frac{\left\langle \epsilon^{2}\right\rangle _{0}\beta_{f}^{2}}{n_{\mathrm{max}}}\right\} =\mathcal{O}\left\{ \left(\beta_{f}\Lambda\right)^{2}\right\}, \label{eq:order_N}
\end{equation}
which is small by definition. Therefore, we should not expect to remain in this low-$\left|\beta\right|$
regime long enough for this epoch to significantly contribute to the total number of e-foldings.

Once $-\beta$ becomes comparable to $\Lambda$ it is harder to make predictions as the specific shape of $g\left(\epsilon\right)$
we are working with starts to make a difference. In other words, as $-\beta$ increases,
we start needing more and more higher-order terms in the expansion in Eq.~\ref{eq:mu_T_inf}
which makes model-independent predictions impossible. Nevertheless, we know
$\beta$ will have to keep evolving towards $-\infty$ and, sooner or later,
we will be in the opposite limit where $-T\ll\Lambda$ and we can make use of the fact that
holes should behave like either matter or radiation.

If holes behave like matter then
\begin{multline}
N=\frac{1}{3}\intop_{\rho_{i}}^{\rho_{f}}\frac{d\rho}{\rho_{\mathrm{max}}-\rho}
=-\frac{1}{3}\intop_{\rho_{\mathrm{max}}-\rho_{i}}^{\rho_{\mathrm{max}}-\rho_{f}}\frac{dx}{x}
\\=\frac{1}{3}\ln\left[\frac{\rho_{\mathrm{max}}-\rho_{i}}{\rho_{\mathrm{max}}-\rho_{f}}\right]=\frac{1}{3}\ln\left[\frac{1+w_{i}^{-1}}{1+w_{f}^{-1}}\right]\label{eq:DN_matterhole}
\end{multline}
where $w_{i}$ and $w_{f}$ are the initial and final $w$, respectively.

If instead holes behave like radiation then
\begin{multline}
N=\frac{1}{4}\intop_{\rho_{i}}^{\rho_{f}}\frac{d\rho}{\rho_{\mathrm{max}}-\rho}
=\frac{1}{4}\ln\left[\frac{\rho_{\mathrm{max}}-\rho_{i}}{\rho_{\mathrm{max}}-\rho_{f}}\right]
\\=\frac{1}{4}\ln\left[\frac{\left(1+w_{i}\right)\left(1-3w_{f}\right)}{\left(1+w_{f}\right)\left(1-3w_{i}\right)}\right]\label{eq:DN_radiationhole}
\end{multline}
with essentially the same type of behaviour.

Notice that the density becomes exponentially close to $\rho_{\rm{max}}$ in just a few e-foldings, since Eq.~\eqref{eq:DN_radiationhole} implies that
\begin{equation}
\rho_f  = \rhomax - (\rhomax-\rho_i)e^{-4N}
\end{equation}
and $\rho_i = \mathcal{O}(\rhomax/2)$. An analogous result holds for Eq.~\eqref{eq:DN_matterhole}).

In addition, note that if we compute the adiabatic sound speed
\begin{equation}
c_{s}^{2}\equiv \frac{\dot{P}}{\dot{\rho}},\label{eq:sound_speed}
\end{equation}
we have
\begin{equation}
c_{s}^{2}=\begin{cases}
1+4\frac{\ln2}{\beta^{2}}\frac{n_{\mathrm{max}}}{\left\langle \epsilon^{2}\right\rangle _{0}}=\mathcal{O}\left\{ \frac{1}{\left(\beta\Lambda\right)^{2}}\right\} \gg1 & \text{if}\ \left|\beta\Lambda\right|\ll1\\
0  \qquad\text{if holes behave like matter}\\
\frac{1}{3}  \qquad\text{if holes behave like radiation}
\end{cases}\label{eq:sound_speed_limits}
\end{equation}
which shows that the sound speed only seems to be problematically
large in the very high (negative) temperature regime which should only
be valid at most during a very short time interval.

\subsection{NATive bouncing Universe\label{sub:NATive-bounces}}

Let us now turn our attention to a scenario where the Universe is
contracting (i.e. $H<0$) and, normally, there would be a Big Crunch.
For simplicity, we shall assume a spatially flat Universe (in the end
we should expect the same type of qualitative evolution).

With an energy cut-off, a fermion component cannot be indefinitely
compressed due to the Pauli exclusion principle. So either the fermions have to be destroyed
as the universe collapses, or the contraction has to stop, preventing a Big Crunch (or there is new physics).
If $w=-1$ exactly,
so that $\rho=\rhomax$ and contraction does not change the temperon energy density,
we have the situation where fermions are destroyed at just the right rate for exponential contraction to continue indefinitely.
However, in other cases we can hope for a bounce.

An expanding Universe tends towards $T=0^{\pm}$ (depending
on the initial sign of $T$), but in the contracting case it should tend
towards $\beta=0$%
\footnote{Note that an interesting consequence of this fact is that the mere
existence of the energy cut-off will lead to exotic cosmological dynamics
due to ``excess'' positive pressure (in particular, as we shall
see, possibly preventing a Big Crunch) even if the Universe
is at a positive temperature all the time.%
}.
This is because the energy conservation equation forces $\dot{\rho}$
to have the same sign as $T$ and to be proportional to $-\frac{H}{\beta}$ once $\left|\beta\right|$
becomes sufficiently small. This causes $\rho$
to approach $\frac{1}{2}\rho_{\mathrm{max}}$, corresponding to $\beta=0$
(recall that $\rho+P$ must change sign at that point).
At some point, then, the small $\beta$ approximation must become
valid and we can follow the evolution of $H$ analytically%
\footnote{If one simply wishes to verify it is not possible to contract forever
in this regime it suffices to take a look at Eq.~\ref{eq:DN_T_inf}
(for which the sign of $H$ makes no difference) and confirm that
$N$ is bounded.%
}.
Note also that the dynamics of this system should not
change appreciably even in the presence of other (normal)
types of matter. This is because the NATive pressure singularity (which occurs for finite $a$) should dominate
the Friedmann equations even if the energy density of temperons is subdominant
(as for "normal" matter $\rho + P$ can only diverge when $a=0$).

Combining Eqs.~\ref{eq:Friedmann_eq_Lambda0_k0} and \ref{eq:beta_zero_rho},
we can find a relation for the temperature as a function of $H^{2}$
\begin{equation}
\beta=\frac{2\rho_{\mathrm{max}}-12H^{2}}{\left\langle \epsilon^{2}\right\rangle _{0}}.\label{eq:beta_H_smallbeta}
\end{equation}
Using this we can write
\begin{equation}
\frac{dH}{dt}=-\frac{\ln2}{2}n_{\mathrm{max}}\frac{\left\langle \epsilon^{2}\right\rangle _{0}}{2\rho_{\mathrm{max}}-12H^{2}},\label{eq:dH_smallbeta}
\end{equation}
which can be integrated to yield
\begin{multline}
-2\left(H^{3}-H_{i}^{3}\right)+\rho_{\mathrm{max}}\left(H-H_{i}\right)+\frac{\ln2}{4}n_{\mathrm{max}}\left\langle \epsilon^{2}\right\rangle _{0}\left(t-t_{i}\right)
\\=0.\label{eq:3rdorder_smallbeta}
\end{multline}
This encodes the evolution of $H\left(t\right)$ in a cubic equation; it has a well-known set of solutions, however it is easier
to understand what happens next graphically.

\begin{figure}[h]
\includegraphics[width=10cm]{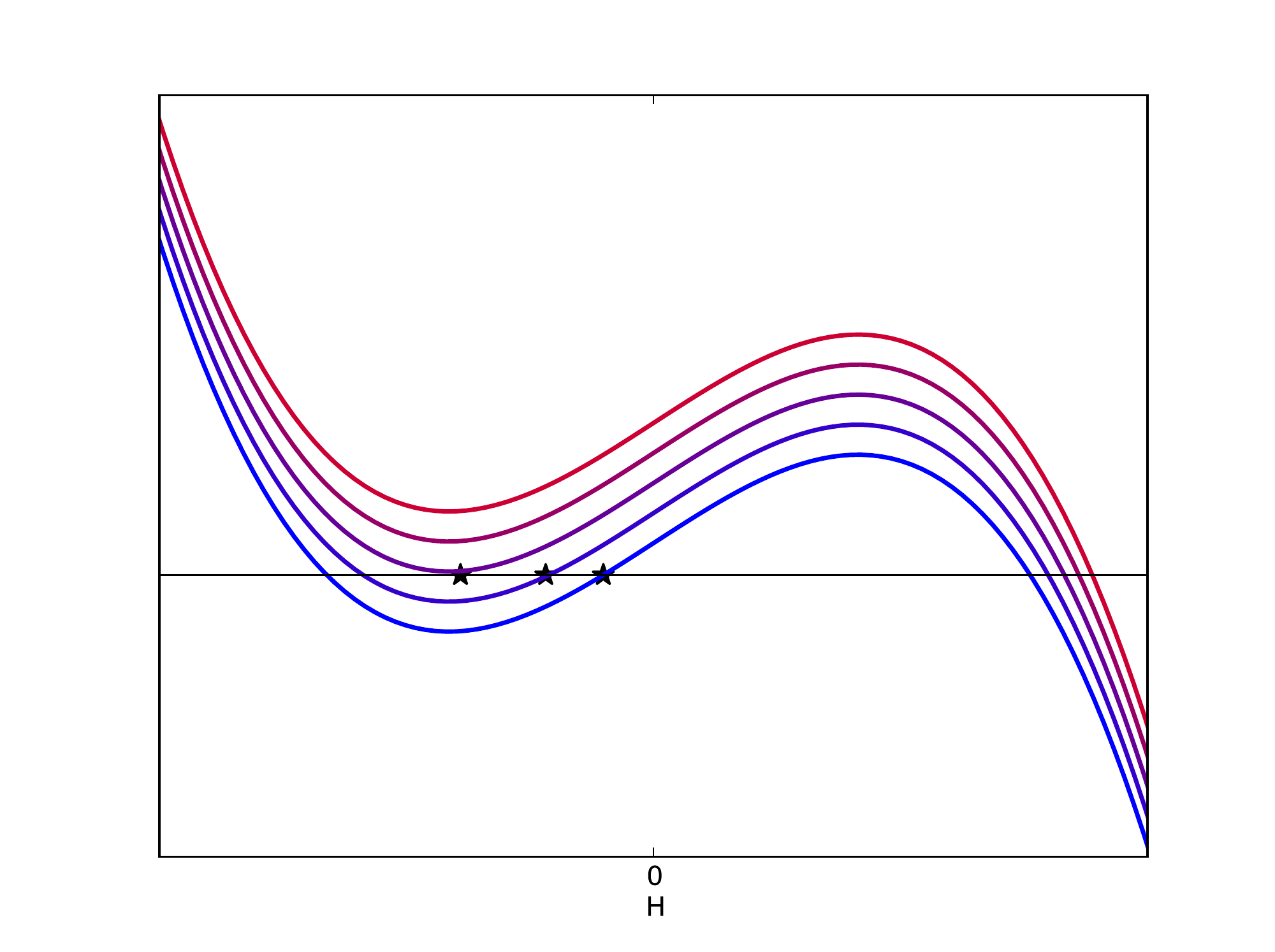}

\caption{A graphic representation of Eq.~\ref{eq:3rdorder_smallbeta}
for increasing values of $t$ starting from $t_{i}$ (increasing from
blue to red and bottom to top). The physical value of $H$ (when it exists)
is indicated by a black star. The proportions between $\left\langle \epsilon^{2}\right\rangle _{0}$,
$n_{\mathrm{max}}$, and $\rho_{\mathrm{max}}$ here correspond to those in Eqs.~\ref{eq:n_rho_eps2_max},
however, it can be shown that a different scenario would look qualitatively
the same.
\label{fig:polynomial_H}}
\end{figure}

From Eq.~\ref{eq:3rdorder_smallbeta} we can see that $H$ at
a given time is given by a root of a third order polynomial whose
zeroth order coefficient is proportional to $t-t_{i}$ (see Fig.~\ref{fig:polynomial_H}).
At time $t=t_{i}$ there are three such roots, the physical solution
corresponding to the middle one ($H=H_{i}$), which must be followed
by continuity until the moment the temperature (and thus $\dot{H}$)
becomes infinite (when $H^{2}=\frac{1}{6}\rho_{\mathrm{max}}$, meaning $\beta=0$).
At that point, the root we were following disappears and there is
no physically meaningful solution to Eq.~\ref{eq:3rdorder_smallbeta}
\footnote{Note that we are not entitled to then follow the remaining root, as
it always corresponds (at this time) to $\rho=2\rho_{\mathrm{max}}$, which
is clearly physically impossible.%
} --- which is not surprising since our formula for the pressure yields a division by zero at this point.
Given that our equations are clearly invalid, we have to resort to physical arguments
in order to know what must happen next.
If we impose that the energy density is continuous and the thermal
equilibrium assumption remains valid then $H$ must
change sign discontinuously causing a bounce.
However, since the pressure is discontinuous
at that point, this is still not enough to determine the subsequent
cosmological evolution. Both a scenario with $\dot{\beta}<0$
 leading to the kind of NATive inflation discussed in subsection \ref{sub:NATive-inflation}
and a scenario with $\dot{\beta}>0$ leading immediately to a ``normal''
expanding Universe seem possible.
The discontinuity in $H$ is likely to be an indication that
our approach is not valid at the moment of the bounce.
Nevertheless, it is not unreasonable
to assume that thermal equilibrium should be restored relatively quickly
after the bounce, leading to one of these two options.

As in the previous section, a contracting Universe can solve the horizon problem.
 In this case, the mechanism would be essentially the same as in most other bouncing Universe models:
 homogeneity would be brought about by a large positive pressure
acting during a cosmological contraction.
In order to solve this
problem, bouncing cosmology models need to allow the quantity
\begin{equation}
\mathcal{N}_{H}\equiv\ln\left|aH\right|\label{eq:ekpyrefolds}
\end{equation}
to grow by a factor of order $60$ \cite{run_bounce}. This seems
to be achieved as long as the contraction starts at sufficiently small
$H$. For example, assuming a matter or radiation dominated Universe
at the beginning of the contraction,
\begin{equation}
\mathcal{N}_{H}\left(t\right)-\mathcal{N}_{H}\left(t_{i}\right)=\left(\frac{2}{3\left(1+w_{0}\right)}-1\right)\ln\left|\frac{H_{i}}{H\left(t\right)}\right|\label{eq:efolds_w_ekpyrosis}
\end{equation}
where $w_{0}=1/3$ if radiation dominates.
Since the left-hand side of Eq.~\ref{eq:efolds_w_ekpyrosis}
is always negative and $H^{2}$ is increasing during the contraction,
it is always possible to get the right amount of contraction as long
as the initial energy density is low enough%
\footnote{Actually, one might raise the question of whether we are demanding
the initial energy density to be too low. Assuming that $\mathcal{N}\left(t_{\star}\right)-\mathcal{N}\left(t_{i}\right)\sim60$
and that $H\left(t_{\star}\right)$ is low enough that we can still
treat temperons as radiation, as is implicit in Eq.~\ref{eq:efolds_w_ekpyrosis},
then $H_{i}/H_{\star}\sim e^{-120}$. In a flat Universe
this would correspond to $\rho_{i}/\rho_{\star}\sim10^{-104}$,
which is not a particularly small number if we keep in mind that if
$\rho_{\mathrm{max}}$ is of order $10^{111}\mathrm{GeVm^{-3}}$ (the maximum
order of magnitude for $\rho$ during inflation from tensor modes
constraints) then the ratio between the critical energy density today
and $\rho_{\mathrm{max}}$ is $\sim10^{-110}$. Moreover, Eq.~\ref{eq:efolds_w_ekpyrosis}
should underestimate ${\cal N}$ since at very late times
a correct formula should account for the diverging increase in positive
pressure.%
}.

If, instead, the Universe is initially at a very low negative temperature
(let us assume, for simplicity, that holes behave like radiation), then
\begin{equation}
\mathcal{N}_{H}\left(t\right)-\mathcal{N}_{H}\left(t_{i}\right)=\frac{1}{4}\ln\left|\frac{\rho^{2}\left(t\right)\left(\rho_{\rm{max}}-\rho\left(t\right)\right)}{\rho^{2}_{i}\left(\rho_{\rm{max}}-\rho_{i}\right)}\right|\label{eq:efolds_nat_ekpyrosis}
\end{equation}
which can also be as large as needed provided that the initial hole density is small enough.

This would mean the NAT themselves would not really contribute to
solving the horizon problem (though the extra positive
pressure close to the bounce would help). In fact, NAT might not even
occur in this scenario --- it is enough for temperons to force a bounce
in a model that would otherwise still solve these problems but end
in a Big Crunch.

\subsection{Perturbations\label{sub:A-NATive-Curvaton}}

If NATive models are to be taken as realistic candidates to realise
inflation or bouncing cosmologies, then a complete study of perturbation
generation will be necessary. One of the main successes of standard
inflation is how easy it is to write down a model which yields a nearly
scale-invariant spectrum of scalar perturbations (which is in excellent
agreement with CMB observations).
One might think that the nearly-exponential expansion of a NAT fluid also would give a
scale invariant spectrum of thermal fluctuations. However an exactly de Sitter phase produces no density fluctuations,
and this limit is rapidly approached. Moreover, $\rho+P$ goes to zero sufficiently fast that
curvature perturbations rapidly increase with time, leading to a blue spectrum
until the de Sitter limit is saturated (see Appendix \ref{sec:Thermal-Perturbation-Generation} for details).
Thus pure NATive inflation cannot be a realistic model for the early Universe.

Instead we can consider a simple scenario with a spectator field that has negligible effect on the background evolution.
Suppose that besides the temperons
there exists a canonical scalar field $\sigma$ whose potential $V\left(\sigma\right)$
is much smaller than the temperon energy density (and does not interact with temperons).
Since the background evolution is almost unchanged, the temperon density will still tend
towards its maximum possible value with  $P_{t}=-\rho_{t}=-\rho_{\mathrm{max}}$,
and its contributions to the Friedmann equations will quickly become
constant. The evolution is then the same as we would have for
just a canonical scalar field with potential $V_{\rm eff}\left(\sigma\right)=V\left(\sigma\right)+\rho_{\mathrm{max}}$.

We can also look at the curvature perturbation produced in the same limit.
Using the fact that the density
perturbation due to temperons should tend to zero (since $\delta\rho_{t}=-\delta\rho_{\rm holes}$
and there are no holes at $T=0$), we have the interesting result that in the flat slicing
\begin{equation}
\zeta=-H\frac{\delta\rho}{\dot{\rho}}\rightarrow -H\frac{\delta\rho_{\sigma}}{\dot{\rho}_{\sigma}}\equiv\zeta_{\sigma},\label{eq:zeta_def-1}
\end{equation}
where $\zeta_{\sigma}$ is the curvature perturbation we would get
from the same scalar field (with potential $V_{\rm eff}\left(\sigma\right)=V\left(\sigma\right)+\rho_{\mathrm{max}}$).
However, since the spectator field has (by construction) negligible density, this would not significantly contribute
to an observable curvature perturbation if the dominant uniform temperon density somehow decays to give a radiation dominated universe. Instead,  the spectator field fluctuation would either have to become dynamically important after temperon decay or
somehow modulate the decay process. We discuss this further at the end of the next section.

\subsection{Ending NATive inflation\label{sub:The-end-of-NAT}}

The analysis so far has focused mostly on the basic cosmological implications
of the possibility of domination by a temperon gas. Since an inflationary
and a bouncing Universe both seem to be naturally realized in this
sort of scenario, it is worth considering whether the transition from NATive inflation to
a normal positive-temperature Universe --- which we may call \emph{recooling},
by analogy with reheating --- can also happen naturally, giving rise
to the standard Hot Big Bang cosmology.

The main difficulty in an expanding universe is that we have shown that a NAT fluid rapidly tends to the stable attractor solution with constant density $\rho=\rhomax$, so on its own there is no dynamical evolution that could naturally set a timescale for recooling. However, as with reheating, the process of recooling to a universe dominated by familiar content must require some level of interaction with normal particles, however indirect, so it is possible that additional degrees of freedom could be responsible for ending NATive inflation.

Note that for $\rho > \half\rhomax$, the energy conservation equation for the temperons has $\dot{\rho_t} = -3H(\rho_t+P_t)  > 0$ (with singular negative pressure term at the $\rho=\half\rhomax$ threshold between positive and negative temperatures), which prevents the temperon fluid from dynamically evolving to normal temperatures even if other components modify the background. The temperons also cannot be in equilibrium with normal matter (involving particles which do not admit negative temperatures), since equilibrium would be reached with both systems at a positive temperature,
regardless of how small the additional positive-temperature system might be \cite{NO_NEGT}. Any end to the NATive epoch must therefore involve an out of equilibrium process.

With this in mind, if we naively postulate that above a critical energy density
temperons can interact with bosons slightly and even decay into bosons
with some low probability, we should expect to recover a positive-temperature
Universe some time after that critical energy density is reached.
This whole process would necessarily take us away from
equilibrium, so the formalism we have been using is no longer
valid and it is not possible to
make model-independent predictions. It seems plausible that it should
be possible to get more e-foldings of inflation by forcing the
temperon-photon interaction to be weaker, at the possible expense of fine tuning the interaction timescale to be
close to the Hubble time.
However we can see from
Eqs.~\ref{eq:DN_matterhole} and \ref{eq:DN_radiationhole} that we would not get
more than a few e-foldings of expansion in equilibrium
unless the critical energy density is also fine-tuned to be extremely close
to $\rho_{\mathrm{max}}$: if we wanted about $60$ e-foldings in
this regime we would need $1-\frac{\rho_{c}}{\rho_{\mathrm{max}}}\lesssim10^{-6}$.
Note also that out of equilibrium
the perturbation calculations from Appendix \ref{sec:Thermal-Perturbation-Generation}
would also not be applicable.

An added difficulty is how to calculate the effective pressure in a non-equilibrium setting.
Unfortunately, this requires calculating the pressure from
first principles, which is non-trivial and model-dependent - even in equilibrium.
The mechanical pressure is usually given by the standard formula
\begin{equation}
P_{\rm{mech}}=\int_m^\Lambda\left(\frac{\epsilon^{2}-m^{2}}{3\epsilon}\right)g\left(\epsilon\right){\cal{N\left(\epsilon\right)}} d\epsilon.
\label{eq:P_standard}
\end{equation}
However, we have been using the result of Eq.~\ref{eq:f(T,mu)} (which assumes thermal equilibrium).
These are not equivalent in the presence of a cutoff, and are only equivalent in the limit
where $\Lambda\rightarrow\infty$ if $\beta>0$. This can be seen
using integration by parts and assuming the ansatz in Eq.~\ref{eq:DOS_Lambda} as well as ${\cal{N}}=1/\left(e^{\beta \epsilon}+1\right)$, which yields
\begin{equation}
P=P_{{\rm mech}}+\frac{g}{2\pi^{2}}\frac{\left(\Lambda^{2}-m^{2}\right)^{3/2}}{3\beta}\ln\left(1+e^{-\beta\Lambda}\right)\label{eq:P_mech_th}
\end{equation}
for $\beta>0$. For negative temperatures the result instead follows Eq.~\eqref{eq:P_holes}.

It may seem a problem that Eq.~\eqref{eq:P_standard} does not work for NAT (and in particular fails to even allow
$P<0$). However, it was originally written down for ideal
gases of classical particles and, at these high energy (and momentum)
scales, close to the cut-off, there is no reason why that picture
should still be valid. It is interesting to note that the pressure in the Friedmann equations
should always coincide with that given by Eq.~\ref{eq:f(T,mu)}.
This can be seen by noting that the first Friedmann equation is equivalent to the First Law of
Thermodynamics in the case of adiabatic expansion/contraction.

Although we do not know how to calculate these "microscopic" pressures, some hints are given by the work of \cite{phantomGUP},
who did something similar for the case of a scalar field, finding that there
was a negative correction to the pressure that made some normal inflation
models become phantom.
The main idea is to make use of the known fact \cite{discrete_heisenberg}
that in these theories there is a significant deviation from the canonical
commutation relation between the usual position and momentum operators,
implying that the usual momentum operator is no longer the conjugate
momentum of the position operator and invalidating the standard result.
In principle, it should be possible to rewrite the Lagrangian in terms
of the correct momentum operator and from that compute the corrections
to the standard energy-momentum tensor due to this deformed algebra.
This would then have the effect of adding corrections to
both the pressure and the energy density%
\footnote{The changes to the energy density being interpretable as differences
in the function $g\left(\epsilon\right)$ due to in one case it being
related to the deformed momentum operator and in another to the actual
eigenvalues of the correct Hamiltonian%
}, essentially solving our problem and enabling the accurate calculation
of non-equilibrium pressures. The pursuit of this approach is left for future work.
In principle, if it succeeds, it may help us understand what is required for
a complete microphysical description of these fluids (at the Lagrangian level). 

An alternative way to end NATive inflation would be to make use of a spectator field as
described in Sec.~\ref{sub:A-NATive-Curvaton}.
A natural way to do this might be for the scalar field to precipitate the end of
inflation, for example by having it decay into bosons
which then interact with the temperons, ending inflation by
full thermalisation. However, since the energy density of the spectator field should be subdominant,
this would probably require a sharp feature in the
potential to compensate the large Hubble damping from the background.

The curvature perturbation from the spectator field (Eq.~\ref{eq:zeta_def-1}) will have a nearly
scale-invariant spectrum during temperon domination provided that it is light compared to the Hubble scale,
but it does not automatically give rise to a significant amplitude of the curvature perturbation after recooling. Note that with two independent components
$\zeta$ can change in time on superhorizon scales \cite{WANDS_SEPARATE}.
However, the temperon fluid should be very homogeneous, so the
recooling surface would be determined by the scalar
field perturbations $\delta\rho_{\sigma}$. The quasi-scale-invariant $\delta\rho_\sigma$ fluctuations can therefore convert into
local variation in the recooling time, and hence a total curvature perturbation (i.e.~essentially the same mechanism as the modulated
reheating mechanism for multi-field inflation \cite{inhom_reheating_dvali}).
Non-Gaussianities could also be introduced at this stage
analogously to similar scenarios in the context of inflation \cite{reviewNG_multif,local_NG_inflation}.
A specific model would be required to make quantitative predictions, in particular the perturbation amplitude is model dependent even if the fluctuation scale dependence is more generally preserved.

\section{Conclusions\label{sec:Conclusions}}

A fluid with negative temperature is an interesting effective macroscopic model for a component of the Universe, even without any compelling microphysical motivation.
Regardless of the microphysics, the evolution of a NAT fluid dominated cosmology will be qualitatively the
same and depend only on the initial value of the Hubble parameter
(as summarised in table \ref{tab:Fate-of-a-NATive-uni}). It might be an attractive way to realise both inflation and bouncing cosmologies.

\begin{table}[h]
\begin{centering}
\begin{tabular}{|c|c|c|}
\hline
 & $H<0$ & $H>0$\tabularnewline
\hline
\hline
$H^{2}<\frac{1}{6}\rho_{\mathrm{max}}$ & NATive bouncing & standard cosmology\tabularnewline
\hline
$H^{2}>\frac{1}{6}\rho_{\mathrm{max}}$ & NATive bouncing & NATive inflation\tabularnewline
\hline
\end{tabular}
\par\end{centering}

\caption{Fate of a temperon-dominated Universe depending on its initial conditions.\label{tab:Fate-of-a-NATive-uni}}
\end{table}

However, there are a number of significant problems with a naive application to cosmology:
\begin{itemize}
\item The physical plausibility of obtaining a maximum energy cutoff is unclear.
\item For a NAT description to apply, the system must remain in equilibrium. In an expanding universe the NAT fluid has to be able to produce more particles rapidly as the universe expands (but no bosons), and any microphysical model would have to explain why this happens rather than simply rapidly decoupling and going out of equilibrium.
\item In an expanding universe, the NAT component rapidly becomes indistinguishable from a cosmological constant at the background level, and it is therefore of limited interest for obtaining realistic dynamics that lead to the end of inflation.
\item An additional component, such as a light scalar spectator field, would be required to produce an acceptable fluctuation spectrum.
\item Any end of NATive inflation, or resolution of a bounce, requires non-equilibrium evolution that cannot be modelled in a model-independent way.\newline
\end{itemize}

Nevertheless, it must be noted that, as long as the first assumption (about the existence
of the energy cut off) holds, there will be interesting consequences even if the following problems cannot be overcome
and our formalism cannot be used most of the time. Even if thermalization at a NAT turns out to
be impossible, a universe with a cut-off would likely still lead to interesting dynamics (for example due to the pressure discontinuity at $T=+\infty$).
Moreover, even if this component turns out to be unable to provide an acceptable explanation to
horizon and flatness problems, it may still have interesting consequences in systems where it might be found
if it exists --- as in the interior of black holes.

The discussion of Appendix \ref{sec:Thermal-Perturbation-Generation} suggests some ways in which acceptable
fluctuations could be produced, though with additional ingredients and fine tuning such models have limited appeal.

Above all, we have shown that there are interesting cosmological consequences
of NAT, and that it is possible that popular paradigms like inflation
and bouncing cosmologies may be successfully realised in scenarios
which are fundamentally different from the usual domination by simple
scalar fields.

\section{Acknowledgements}

We thank Mariusz D\k{a}browski for bringing the work of \cite{Phantom_thermo}
to our attention and Yi Wang for helpful comments on \cite{thermo_perturb}.
We also thank Carlos Martins, Daniel Passos, Daniela Saadeh, Frank K\"{o}nnig, Henryk
Nersisyan, Jonathan Braden, and Sam Young for fruitful discussions, insightful
questions, and other useful references.

JV is supported by an STFC studentship, CB is supported by a Royal
Society University Research Fellowship, and AL acknowledges support
from the Science and Technology Facilities Council {[}grant number
ST/L000652/1{]} and European Research Council under the European Union's Seventh Framework Programme (FP/2007-2013) / ERC Grant Agreement No. [616170].

\bibliography{PhD_BIB}

\appendix

\section{Problems maintaining negative temperatures with number conservation and bosons}
\label{sec:The-problem-of-mu}

Let us start by assuming number conservation so that we can explore the kind of problems it causes.
In an FLRW Universe with
scale factor $a$ and Hubble parameter $H$, there will then be two
equations governing the dynamics of these functions,
the continuity equation
\begin{equation}
\dot{\rho}=-3H\left(\rho+P\right)\label{eq:continuity_FLRW}
\end{equation}
and the number conservation equation
\begin{equation}
\frac{\dot{n}}{n}=-3H.\label{eq:n_conservation}
\end{equation}
Equivalently, one can make use of the symmetries in Eqs.~\ref{eq:rho_holes}, \ref{eq:n_holes}, and \ref{eq:P_holes}
to rewrite these for holes as
\begin{equation}
\dot{\rho_{h}}=-3H\left(\rho_{h}+P_{h}\right)\label{eq:continuity_FLRW_holes}
\end{equation}
\begin{equation}
\dot{n_{h}}=3H\left(n_{\rm{max}}-n_{h}\right) \label{eq:n_conservation_holes}
\end{equation}
where the subscript $h$ denotes a quantity relative to holes (with $T_{h}=-T$).
Note that the formal equivalence between Eqs.~\ref{eq:continuity_FLRW} and \ref{eq:continuity_FLRW_holes}
is due to the fact that they both express the constraint that the entropy be conserved
and entropy cannot distinguish between particles and holes (since it is purely combinatorial).

Suppose now that the Universe is filled with a temperon gas at NAT with $m\gg T_{h}$. If
the very low-energy holes behave like normal (pressureless) matter then $\rho_{h}=m n_{h}$
and the previous equations are reduced to
\begin{equation}
\dot{\rho_{h}}=-3H\rho_{h}\label{eq:continuity_FLRW_holes_bad}
\end{equation}
\begin{equation}
\dot{\rho_{h}}=3Hm n_{\rm{max}}-3H\rho_{h} \label{eq:n_conservation_holes_bad}
\end{equation}
which is an inconsistent system as long as $H\neq 0$. This would mean
that in this situation equilibrium could not be maintained during the expansion.

Of course, this problem assumes a specific low-energy form of $g\left(\epsilon\right)$ and
thus it can by no means be considered a refutation of the $\mu \neq 0$ case.
Nevertheless, it is a difficulty that has to be taken into account and which raises questions about
how model-independent (i.e., how independent of $g\left(\epsilon\right)$) such an analysis can be.
In addition, if we just assume $g\left(\epsilon\right)$ is whatever is necessary
to make this system of equations consistent, we have to live with the fact
that there are possible situations in which the energy density will be increasing
while the number density decreases (and vice-versa, if the Universe is contracting),
since $\dot{\rho}/\dot{n}$ has the same sign as $\rho +P$ and $\rho$ is bounded
whereas $P$ is not.

Moreover, it can be easily seen that, even accepting these odd behaviours,
such a solution can never be consistent in all situations.
For example, consider the case where temperons are massless fermions at $T=0^{-}$.
In this situation, the energy density must be constant and equal to its maximum possible value,
whereas the number density must vary according to $H$. This is clearly absurd
as there is no way the system can be at $\rho = \rho_{\rm{max}}$ unless all states are filled.

Note that once we restrict ourselves to the study of cases without number conservation
it becomes clear that we cannot use bosons: without number conservation,
the energy of the system is no longer bounded from above, which makes
NAT impossible.

\section{Thermal Perturbation Generation\label{sec:Thermal-Perturbation-Generation}}

A system in equilibrium will in general have thermal fluctuations.
Here we consider the case where the Universe is dominated by temperons in thermal equilibrium,
and calculate the density and curvature perturbations produced. We focus on the case where holes behave
like radiation as $\beta\rightarrow-\infty$%
\footnote{Ignoring any contributions from whatever process might be responsible
for ending inflation.%
}.

The main difference between perturbations here and in the standard
inflationary scenario is that density perturbations here are produced
due to classical thermal fluctuations rather than by quantum effects.
The basic methodology used in this subsection is therefore essentially
the same as the one used to work out thermal fluctuations in models
such as chain inflation or warm inflation in the very weakly dissipative
regime \cite{thermo_perturb}.

\subsection{Moments in position space}

For a canonical thermal system with volume $V$, the $n$-th moment of the energy density
distribution is given by
\begin{equation}
\left\langle \rho^{n}\right\rangle =\frac{\left\langle E^{n}\right\rangle }{V^{n}}=\frac{1}{Z}\left(-\frac{1}{V}\right)^{n}\frac{\partial^{n}Z}{\partial\beta^{n}}\label{eq:moment_density}
\end{equation}
where $Z$ is the partition function as given by Eq.~\ref{eq:Zg_fermions}
with $\mu=0$. Making the substitution $\partial_{\alpha}\equiv-V^{-1}\partial_{\beta}$,
we can find the simple recursive relation
\begin{equation}
\left\langle \rho^{n+1}\right\rangle =\left[\left\langle \rho\right\rangle +\partial_{\alpha}\right]\left\langle \rho^{n}\right\rangle \label{eq:recursive_moment_density}
\end{equation}
which we can then use to find an analogous formula for the moments
of $\delta\rho=\rho-\left\langle \rho\right\rangle $. Taking a derivative
with respect to $\alpha$ of
\begin{equation}
\left\langle \left(\delta\rho\right)^{n}\right\rangle =\sum_{k=0}^{n}\frac{n!}{\left(n-k\right)!k!}\left(-\left\langle \rho\right\rangle \right)^{n-k}\left\langle \rho^{k}\right\rangle \label{eq:moment_deltarho}
\end{equation}
and then making use of Eq.~\ref{eq:recursive_moment_density} yields

\begin{equation}
\left\langle \left(\delta\rho\right)^{n+1}\right\rangle =\partial_{\alpha}\left\langle \left(\delta\rho\right)^{n}\right\rangle +n\partial_{\alpha}\left\langle \rho\right\rangle \left\langle \left(\delta\rho\right)^{n-1}\right\rangle .\label{eq:recursive_moment_deltarho}
\end{equation}

Eq.~\ref{eq:recursive_moment_deltarho} has the interesting feature
of separating contributions from even and odd momenta in the right-hand
side. Because of it, and since $\left\langle \delta\rho\right\rangle =0$,
if we assume that $\left\langle \left(\delta\rho\right)^{3}\right\rangle =\partial_{\alpha}^{2}\left\langle \rho\right\rangle =0$
then (as can be easily checked by induction) every odd moment has
to be zero and $\left\langle \left(\delta\rho\right)^{2n}\right\rangle =\left(2n-1\right)!!\left\langle \left(\delta\rho\right)^{2}\right\rangle ^{n}$,
corresponding to exactly Gaussian perturbations whose statistics depend
only on the size of the thermal system $V$ (and not on $\beta$).
In other words, in this scenario Gaussianity is equivalent to the
third moment of $\delta\rho$ being null and to the second (and indeed
every even) moment being independent of $\beta$. Since
\begin{equation}
\left\langle \left(\delta\rho\right)^{3}\right\rangle =\partial_{\alpha}^{2}\left\langle \rho\right\rangle =\frac{1}{V^{2}}\intop g\left(\epsilon\right)\epsilon^{3}\left[\frac{e^{2\beta\epsilon}-e^{\beta\epsilon}}{\left(e^{\beta\epsilon}+1\right)^{3}}\right]d\epsilon,\label{eq:3rd_moment}
\end{equation}
then density perturbations must always be exactly Gaussian for $\beta=0$
(at the bounce) and for the attractors $\beta=\pm\infty$. This exact
Gaussianity at the attractors, however, is misleading since it corresponds
to a limit in which density perturbations must vanish --- recall that
from Eq.~\ref{eq:rho_holes} we have $\delta\rho=-\delta\rho_{\rm holes}$
(where the temperature of holes is symmetric to that of particles)
and $\delta\rho$ has to be zero for $T=0^{+}$ since at that temperature
the density itself is zero. We should thus look at the relevant perturbation,
the curvature perturbation, which can be written in terms of the density
perturbation in the zero curvature frame (in which the previous
calculations make sense as the shape of the ``box'' is not perturbed)
as
\begin{equation}
\zeta=\frac{1}{3}\frac{\delta\rho}{\left\langle \rho\right\rangle +\left\langle P\right\rangle }.\label{eq:zeta_def}
\end{equation}
In the case of a bounce, this also shows that even the Gaussian perturbations
are less interesting than one might think, since despite
the numerator being non-zero the denominator diverges, making the
relevant curvature perturbation negligible.

The variance of the curvature perturbation can now be found to be
given by
\begin{equation}
\left\langle \zeta^{2}\right\rangle =\frac{1}{9}\frac{\left\langle \delta\rho^{2}\right\rangle }{\left(\rho+P\right)^{2}}=-\frac{1}{9V}\frac{\partial_{\beta}\rho}{\left(\rho+P\right)^{2}},\label{eq:variance_form}
\end{equation}
where for simplicity we are using $\rho$ and $P$ interchangeably
with their averages.

The most interesting limit for Eq.~\ref{eq:variance_form} is when
$\beta\rightarrow-\infty$ as the spacetime then tends towards unperturbed
de Sitter, yet the curvature perturbation does not necessarily tend
to zero as the denominator in the right-hand side also goes to zero
in this limit%
\footnote{These calculations may not even be physically
meaningful too close to that limit, since then most Hubble volumes
will have no holes and will be indistinguishable from de Sitter space,
for which $\zeta$ is not well defined since there isn't a unique
constant-density frame.
There could also be additional effects, for example if the equilibration time is not
negligibly smaller than the Hubble time the density perturbation could be dominated by
fluctuations in the equilibration process.
}. For example, if holes behave like radiation, the denominator vanishes
at a faster rate than the numerator, causing $\zeta$ to diverge as
(see table \ref{tab:Summary-of-relevant-rad} for a breakdown of the
relevant terms in Eq.~\ref{eq:variance_form} in this limit)
\begin{equation}
\left\langle \zeta^{2}\right\rangle =\left(\frac{15}{7\pi^{2}g}\right)^{1/4}\frac{\left(\rho_{\mathrm{max}}-\rho\right)^{-3/4}}{2V}.\label{eq:variance_zeta}
\end{equation}

\begin{table}[h]
\begin{centering}
\begin{tabular}{|c|c|c|}
\hline
 & $\rho_{h}=\rho_{\mathrm{max}}-\rho$ & $a_{\star}=aV_{0}^{-1/3}$\tabularnewline
\hline
\hline
$\rho+P$ & $-\frac{4}{3}\rho_{h}$ & $-\frac{4}{3}a_{\star}^{-4}$\tabularnewline
\hline
$\beta$ & -$\left(\frac{7\pi^{2}}{240}\frac{g}{\rho_{h}}\right)^{1/4}$ & $-\left(\frac{7\pi^{2}}{240}g\right)^{1/4}a_{\star}$\tabularnewline
\hline
$\left\langle \left(\delta\rho\right)^{2}\right\rangle $ & $\frac{8}{V}\left(\frac{15}{7\pi^{2}g}\right)^{1/4}\rho_{h}^{5/4}$ & $\frac{8}{V}\left(\frac{15}{7\pi^{2}g}\right)^{1/4}a_{\star}^{-5}$\tabularnewline
\hline
$\left\langle \zeta^{2}\right\rangle $ & $\frac{1}{2V}\left(\frac{15}{7\pi^{2}g}\right)^{1/4}\rho_{h}^{-3/4}$ & $\frac{1}{2V}\left(\frac{15}{7\pi^{2}g}\right)^{1/4}a_{\star}^{3}$\tabularnewline
\hline
\end{tabular}
\par\end{centering}

\caption{\label{tab:Summary-of-relevant-rad}Summary of relevant quantities
as functions of hole energy density, $\rho_{h}$, and rescaled scale
factor, $a_{\star}=\rho_{h}^{-1/4}$, when holes behave like radiation.
The row for $\beta$ comes from applying the standard result for the
fermion energy density to $\rho_{h}$. $V_{0}$ is the integration
constant used later in Eq.~\ref{eq:final_P_spectrum}.}
\end{table}

\subsection{The thermal power spectrum}

In order to turn the results from the previous section into predictions
for the power spectrum, we must introduce a couple of mathematical
objects and endure some integral manipulations.

Let us start by considering the average density fluctuation in the
vicinity of a point,
\begin{multline}
\delta\rho_{\mathbf{x_{0}}}\left(r\right)\equiv\frac{1}{V_{r}}\intop_{S_{r,\mathrm{x_{0}}}}d^{3}{\bf x}\delta\rho\left(\mathbf{x}\right)=\frac{1}{V_{r}}\intop_{S_{r}}d^{3}{\bf x}\delta\rho\left(\mathbf{x_{0}+x}\right)
\\=\frac{1}{V_{r}}\intop\frac{d^{3}{\bf k}}{\left(2\pi\right)^{3/2}}\delta\rho_{{\bf k}}\intop_{S_{r}}d^{3}{\bf x}e^{i{\bf k\cdot\left(x_{0}+x\right)}},\label{eq:local_drhor}
\end{multline}
where $S_{r,{\bf x_{0}}}$ is the sphere of comoving radius $r$ centred
around $\mathbf{x_{0}}$, $S_{r}=S_{r,{\bf 0}}$, and $V_{r}=\frac{4}{3}\pi r^{3}$.

We can define the average power of this quantity as
\begin{equation}
\overline{\delta\rho}^{2}\left(r\right)\equiv\lim_{R\rightarrow\infty}\frac{1}{V_{R}}\intop_{S_{R}}d^{3}{\bf x_{0}}\left|\delta\rho_{{\bf x_{0}}}\left(r\right)\right|^{2}\label{eq:deltarhobar}
\end{equation}
which can be rewritten as
\begin{multline}
\overline{\delta\rho}^{2}\left(r\right)=\lim_{R\rightarrow\infty}\intop\frac{d^{3}{\bf k}}{\left(2\pi\right)^{3/2}}\intop\frac{d^{3}{\bf k}^{\prime}}{\left(2\pi\right)^{3/2}}
\\
\times\delta\rho_{{\bf k}}\delta\rho_{{\bf k}^{\prime}}W_{r}\left(k\right)W_{r}\left(k^{\prime}\right)W_{R}\left(\left\Vert {\bf k+k^{\prime}}\right\Vert \right),\label{eq:rhobar_beforeav}
\end{multline}
where we have used the window function defined as
\begin{equation}
W_{r}\left(k\right)\equiv\frac{1}{V_{r}}\intop_{S_{r}}d^{3}{\bf x}e^{i{\bf x\cdot k}}=3\frac{\sin\left(k r\right)-k r \cos\left(k r\right)}{\left(k r\right)^{3}}.\label{eq:window}
\end{equation}

Taking the expected value on both sides and using the definition
$\left\langle \delta\rho_{{\bf k}}\delta\rho_{{\bf k^{\prime}}}\right\rangle \equiv\delta\left({\bf k+k^{\prime}}\right)P_{\delta\rho}\left(k\right)$
then yields
\begin{equation}
\left\langle \overline{\delta\rho}^{2}\left(r\right)\right\rangle =\intop\frac{d^{3}{\bf k}}{\left(2\pi\right)^{3}}\left|W_{r}\left(k\right)\right|^{2}P_{\delta\rho}\left(k\right).\label{eq:bar_spectrum}
\end{equation}
Alternatively, using the usual definition of the power spectrum
${\cal P}\left(k\right)=k^{3}P\left(k\right)/\left(2\pi^{2}\right)$, this is
\begin{equation}
\left\langle \overline{\delta\rho}^{2}\left(r\right)\right\rangle =\intop_{0}^{\infty}\frac{dk}{k}\left|W_{r}\left(k\right)\right|^{2}\mathcal{P}_{\delta\rho}\left(k\right).\label{eq:bar_calspectrum}
\end{equation}

Provided that $P_{\delta\rho}\left(k\right)$ doesn't diverge faster
than $k^{-3}$ as $k\rightarrow0$, the integral in Eq.~\ref{eq:bar_spectrum}
is dominated by $k\sim r^{-1}$ and \cite{mukperturb}
\begin{equation}
P_{\delta\rho}\left(k\right)\sim \frac{\left(2\pi\right)^{3}}{k^{3}}\left\langle \overline{\delta\rho}^{2}\left(k^{-1}\right)\right\rangle .\label{eq:calspectrum_variance}
\end{equation}

Assuming there is a \emph{thermal horizon}%
\footnote{
In the literature, some measure of the typical wavelength of a particle (usually a photon) in the thermal
system has been used as the thermal horizon, although \cite{thermo_perturb}
note it can in principle be any scale between that and the \emph{acoustic
horizon}, $c_{s}H^{-1}$.%
}, $L_{\rm th}$, corresponding to the physical distance beyond which there can
be no thermal correlation, the actual observed density power spectrum
can be calculated from Eq.~\ref{eq:calspectrum_variance} evaluated
around thermal horizon exit, when $k\sim a/L_{\rm th}$. Using $\left\langle \overline{\delta\rho}^{2}\left(L/a\right)\right\rangle =\left\langle \left(\delta\rho\right)^{2}\right\rangle_{V_{\rm th}} $
we conclude that
\begin{equation}
\left\langle \overline{\delta\rho}^{2}\left(r\right)\right\rangle =\left\langle \left(\delta\rho\right)^{2}\right\rangle _{V=a^{3}V_{r}} ,\label{eq:ergodic_rms}
\end{equation}
 and hence
\begin{multline}
P_{\zeta}\left(k\sim a/L_{\rm th}\right)\sim\frac{\left(2\pi\right)^{3} L_{\rm th}^{3}}{9a^{3}\left(\rho+P\right)^{2}}\left\langle \delta\rho^{2}\right\rangle _{V=\frac{4\pi}{3}L_{\rm th}^{3}}\\\sim-\frac{\left(2\pi\right)^{3}}{12 a^{3}}\frac{\partial_{\beta}\rho}{\left(\rho+P\right)^{2}},\label{eq:P_spectrum_general_all}
\end{multline}
where everything is evaluated around thermal horizon crossing.

If we further assume holes behave like radiation, we can use this
together with equation \ref{eq:variance_zeta} and immediately get
\begin{equation}
P_{\zeta}\left(k\right)\sim\frac{3}{8}\left(\frac{15}{7\pi^{6}g}\right)^{1/4}\frac{\left(2\pi\right)^{3}}{V_{0}}, \label{eq:final_P_spectrum}
\end{equation}
where $V_0 \equiv a^{3}\left(\rho_{\mathrm{max}}-\rho\right)^{3/4}$
is a constant thanks to Eq.~\ref{eq:DN_radiationhole}.
Note that this corresponds to
a white noise spectrum with $n_{s}=4$.

If the NAT model were to describe a realistic cosmology, we would need the power spectrum to be (approximately) scale-invariant,
at least in the attractor $\beta\rightarrow-\infty$ limit. Unfortunately,
we can show that is not necessarily possible even if we allow drastic
departures from Eq.~\ref{eq:DOS_Lambda}. From Eq.~\ref{eq:calspectrum_variance}
it is clear that the spectrum will be scale invariant if and only
if $\left\langle \overline{\delta\rho}^{2}\left(k^{-1}\right)\right\rangle $
is independent of $k$. In other words, the power spectrum can be
written as
\begin{equation}
{\cal P}_{\zeta}\left(k\right)\sim-\frac{4\pi}{9V}\frac{\partial_{\beta}\rho}{\left(\rho+P\right)^{2}},\label{eq:power_spectrum_general_thermal}
\end{equation}
which is a constant if and only if
\begin{equation}
\partial_{k}\left(V^{-1}\frac{\partial_{\beta}\rho}{\left(\rho+P\right)^{2}}\right)=0.\label{eq:scale_inv_condition}
\end{equation}
\newline\newline Assuming, as before, that $V\propto L_{\rm th}^{3}=\text{const}$, this can be
rewritten as
\begin{equation}
\frac{\partial_{\beta}\rho^{-1}}{\left(1+w\right)^{2}}=\text{const}.\label{eq:scale_inv_w}
\end{equation}
Note that if $w_{h}$ is the equation of state of holes then Eqs.
\ref{eq:rho_holes} and \ref{eq:P_holes_mu0} yield $\rho+P=-\rho_{h}-P_{h}$
(where the subscript $h$ once again denotes holes) and thus Eq.~\ref{eq:scale_inv_w}
is equivalent to
\begin{equation}
\frac{\partial_{\beta_{h}}\rho_{h}^{-1}}{\left(1+w_{h}\right)^{2}}=\text{const}.\label{eq:scale_inv_holes}
\end{equation}
Consequently, if $w_{h}=\text{const}$ then in this large $\beta>0$ limit
\begin{equation}
\rho=\frac{1}{A+B\beta}, \label{eq:weirdrho}
\end{equation}
 where $A$ and $B$ are positive constants of integration --- they
have to be positive because $B$ is related to the power spectrum
by ${\cal P}_{\zeta}\left(k\right)\sim\frac{4\pi}{9V}\frac{B}{\left(1+w_{h}\right)^{2}}$.
If we now equate the right-hand sides of Eq.~\ref{eq:rho_general}
and Eq.~\ref{eq:weirdrho} and take a derivative with respect to $T$
at $T=0^{+}$ we get
\begin{equation}
\lim_{\beta\rightarrow\infty}\beta^{2}\int\frac{g\left(\epsilon\right)\epsilon^{2}e^{\beta\epsilon}}{\left(e^{\beta\epsilon}+1\right)^{2}}d\epsilon=1\label{eq:absurd}
\end{equation}
which is absurd since the left-hand side should be zero as long as
there is a finite total number of one-particle states.
\end{document}